\documentclass[twocolumn]{aastex62}
\usepackage{graphicx}
\usepackage{color}
\usepackage{amsmath}
\usepackage{comment}
\usepackage{lineno}
\usepackage[first=0,last=9]{lcg}

\usepackage{array,multirow}

\hypersetup{linkcolor=blue,citecolor=blue,filecolor=blue,urlcolor=blue}

\definecolor{amber}{rgb}{1.0, 0.75, 0.0}

\definecolor{amethyst}{rgb}{0.6, 0.4, 0.8}

\definecolor{blue-violet}{rgb}{0.54, 0.17, 0.89}

\definecolor{blue(munsell)}{rgb}{0.0, 0.5, 0.69}

\definecolor{byzantine}{rgb}{0.74, 0.2, 0.64}

\definecolor{carminered}{rgb}{1.0, 0.0, 0.22}

\definecolor{bittersweet}{rgb}{1.0, 0.44, 0.37}

\definecolor{ao(english)}{rgb}{0.0, 0.5, 0.0}

\graphicspath{{./}{figures/}}

\begin{document}

\title{\bf Do gravitational wave observations in the lower mass gap favor a hierarchical triple origin?}
\author{V. Gayathri}%
\affiliation{Leonard E. Parker Center for Gravitation, Cosmology, \& Astrophysics, University of Wisconsin–Milwaukee, Milwaukee, WI 53201, USA}
\affiliation{Department of Physics, University of Florida, Gainesville, FL 32611-8440, USA}
\author{I. Bartos}
\thanks{imrebartos@ufl.edu}
\affiliation{Department of Physics, University of Florida, Gainesville, FL 32611-8440, USA}
\author{S. Rosswog}
\affiliation{Universit\"{a}t Hamburg, D-22761 Hamburg, Germany}
\affiliation{The Oskar Klein Centre, Department of Astronomy, Stockholm University, AlbaNova, SE-106 91 Stockholm, Sweden}
\author{M.C. Miller}
\affiliation{Department of Astronomy and Joint Space-Science Institute, University of Maryland, College Park, MD 20742-2421, USA}
\author{D. Veske}
\affiliation{Institut f\"{u}r Theoretische Physik, Heidelberg University, Philosophenweg 16, 69120 Heidelberg, Germany}
\author{W. Lu}
\affiliation{Department of Astronomy and Theoretical Astrophysics Center, University of California, Berkeley, CA 94720-3411, USA}
\author{S. Marka}
\affiliation{Department of Physics, Columbia University in the City of New York, New York, NY 10027, USA}


\begin{abstract}
Observations of compact objects in Galactic binaries have provided tentative evidence of a dearth of masses in the so-called lower mass gap $\sim2.2-5$\,M$_\odot$. Nevertheless, two such objects have been discovered in gravitational-wave data from LIGO and Virgo. 
Remarkably, the estimated masses of both secondaries in the coalescences GW190814 ($m_2=2.59^{+0.08}_{-0.09}$M$_\odot$) and GW200210\_092254 ($m_2=2.83^{+0.47}_{-0.42}$M$_\odot$) fall near the total mass of $\sim 2.6$\,M$_\odot$ of observed Galactic binary neutron star systems. The more massive components of the two binaries also have similar masses. Here we show that a neutron star merger origin of the lighter components in GW190814 and GW200210\_092254 is favored over $M^{-2.3}$ (Bayes factor $\mathcal{B}\sim 5$) and uniform ($\mathcal{B}\sim 14$) mass distributions in the lower mass gap. We also examine the statistical significance of the similarity between the heavier component masses of GW190814 and GW200210\_092254, and find that a model in which the mass of GW200210\_092254 is drawn from the mass posterior of GW190814 is preferred ($\mathcal{B}\sim 18$) to a model in which its mass is drawn from the overall mass distribution of black holes detected in gravitational wave events.  This hints at a common origin of the primary masses, as well as the secondary masses, in GW190814 and GW200210\_092254.
\vspace{1cm}
\end{abstract}

\section{Introduction} \label{sec:intro}

The latest gravitational wave transient catalog (GWTC-3) of the  LIGO \citep{2015CQGra..32g4001L} and Virgo \citep{2015CQGra..32b4001A} observatories reported $\sim 90$ compact binary mergers, including binary black holes, binary neutron stars, and neutron star-black hole mergers \citep{LIGOScientific:2021djp}. Among these are two remarkable events, GW190814 \citep{2020ApJ...896L..44A} and GW200210\_092254 (\citealt{LIGOScientific2021djp}; although the latter was detected with a false alarm rate of $\sim1$\,yr$^{-1}$), which contain one companion object that falls into the so-called lower mass gap between $2.2\;M_{\odot}$ - $5\;M_{\odot}$. The reconstructed rate density of these events is $1-23$\,Gpc$^{-3}$yr$^{-1}$ \citep{2020ApJ...896L..44A}, only a factor of a few lower than the merger rate density $17.9-44$\,Gpc$^{-3}$yr$^{-1}$ of binary black hole mergers \citep{2021arXiv211103634T}.

Observations of Galactic compact objects have uncovered hints that there may be a paucity of objects in binaries that have masses within the lower mass gap (\citealt{1998ApJ...499..367B,2010ApJ...725.1918O,2011ApJ...741..103F}; but see \citealt{2021ApJ...908L..38A}).  This may be a feature of star formation, although selection effects cannot be ruled out. For example, if core collapse supernovae impart kicks of comparable momentum on newly formed compact objects, this will mean larger kick velocities for lower-mass black holes, which may then be less likely to remain within binaries in which they can be observed.

\begin{figure*}
    \centering
   \includegraphics[width=0.9\textwidth,,trim={0cm 4cm 2.5cm 0cm},clip]{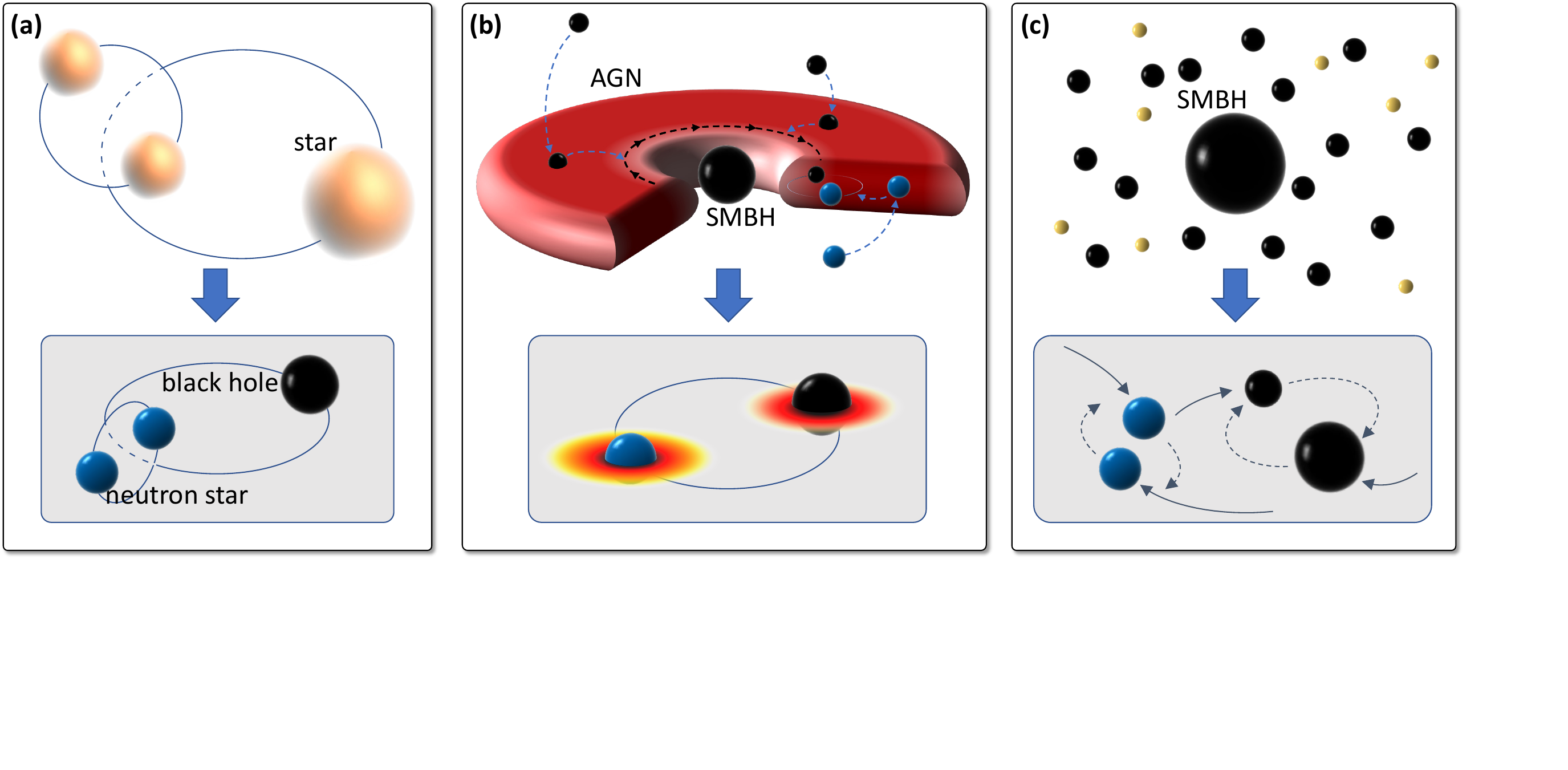}
    \vspace{-0.7cm}
    \caption{{\bf Illustration of possible origins of binary mergers with secondaries in the lower mass gap.} {\bf (a)} Isolated hierarchical triples with stars in the inner binary forming neutron stars that merge into a remnant in the mass-gap. {\bf (b)} Gas capture in AGN disks around a central supermassive black hole (SMBH); the mass gap secondary is a heavily accreting neutron star. {\bf (c)} Stellar clusters produce the dynamical capture of two neutron stars, which merge into a mass gap object; with the remnant consecutively undergoing dynamical capture and merger with a black hole.} 
    \label{fig:illustration}
\end{figure*}

Even if stellar evolution is ruled out, compact objects can still reach a mass within the gap by two means (see Fig. \ref{fig:illustration}). First, a neutron star, formed with mass $\lesssim 2.2$\,M$_\odot$, can grow through gas accretion from its environment. This can occur in a low-mass X-ray binary, although in this case the companion to the resulting mass-gap object is unlikely to become a black hole after gravitational collapse. Gas accretion can also occur in an AGN disk, where the mass-gap object can later merge with a black hole that is also in the disk \citep{2020ApJ...901L..34Y}. 

Second, a neutron star can merge with an object, such as another neutron star. This can occur in dense stellar clusters, where binary-single or binary-binary encounters could result in the eventual coalescence of a neutron star merger remnant with a black hole. However, the expected rate of such consecutive mergers for the observed masses is expected to be low (see e.g. \citealt{2020PhRvD.101l3010S,2020ApJ...888L..10Y,2019MNRAS.482...30S,2020ApJ...895L..15F,2021MNRAS.500.1817L}). In addition, black holes dominate cluster cores and limit the mass segregation of neutron stars \citep{2020ApJ...888L..10Y}. Therefore, the rate density of dynamical capture and merger of black holes should be substantially higher than that of binary neutron star remnants. Given the relatively small difference between the observed rate density of binary black hole mergers and mergers involving an object in the lower mass gap, a dynamical origin appears unlikely. Similarly, AGN-assisted binary mergers should be overwhelmingly black hole -- black hole mergers, with a much smaller rate of neutron star mergers \citep{2020ApJ...901L..34Y}.

Neutron star mergers that can proceed to merge with a black hole could also be part of an isolated, hierarchical triple (or higher-order multiple) system of massive stars, with the two neutron stars forming from the inner binary and the black hole from the tertiary, as first pointed out by \cite{2021MNRAS.500.1817L} and \cite{2021MNRAS.502.2049L}. The expected rate density of double mergers from such hierarchical triples is within $\sim0.1\%-3\%$ of the binary neutron star merger rate density of $10-1700$\,Gpc$^{-3}$\,yr$^{-1}$, as inferred from gravitational wave observations \citep{LIGOScientific2021djp}. It is therefore plausible that a fraction of the observed neutron star mergers eventually also undergoes a secondary merger \citep{2021MNRAS.500.1817L,2023arXiv230210350B}.

The mass of objects in the lower mass gap can help identify their possible origin. In particular, based on the narrow distribution of the total mass of known Galactic binary neutron star systems \citep{1994PhRvL..73.1878F,2012ApJ...757...55O,2013ApJ...778...66K,2019ApJ...876...18F} ($\approx 2.65\pm0.12$\,M$_\odot$, where the error represents 1$\sigma$ uncertainty; \citealt{2019ApJ...876...18F}), we can expect neutron star merger remnants to have a narrow mass distribution with masses around $\sim2.6$\,M$_\odot$, similar to the observed objects \citep{2021MNRAS.500.1817L}. Accretion and stellar evolution, on the other hand, would probably result in a more spread out mass distribution within the gap, because at least several tenths of a solar mass must be accreted by a neutron star to reach this mass. Even if a narrower mass distribution arises from accretion or stellar evolution, there is no known astrophysical reason that this narrow distribution would center around $\sim2.6$\,M$_\odot$. A possible caveat with this model is that if two neutron stars became gravitationally bound due to chance encounters in stellar clusters then they may have not originally resided in a binary neutron star system. The mass distribution of such neutron stars is somewhat broader than for those in binary neutron star systems \citep{2013ApJ...778...66K,2011MNRAS.414.1427V,2011A&A...527A..83Z}.  

For mass-gap objects formed by neutron star mergers, an observable difference between the stellar-cluster and hierarchical-triple origins may be the expected mass distribution of the black hole. This is noteworthy in particular since the primary masses of GW190814 and GW200210\_092254 are similar ($23.2^{+1.1}_{-1.0}$M$_\odot$ and $24.1^{+7.5}_{-4.6}$M$_\odot$, respectively). In the stellar-cluster scenario, the neutron star merger remnant encounters black holes quasi-randomly, resulting in a broad probability density for the primary mass. 

A hierarchical triple scenario might produce a more narrowly distributed primary mass. For example, the probability of forming neutron star--neutron star--black hole triples can be higher for a heavy tertiary black hole \citep{2021MNRAS.500.1817L}. This is because supernova explosions may disrupt a triple system. With a heavy ($\sim20$\,M$_\odot$) tertiary, supernova explosions do not cause the entire system to lose half of its mass, therefore the outer orbit is only weakly perturbed by each supernova. 

Here we probe the origin of the observed objects in the lower mass gap using Bayesian model comparison relying on the companion masses in order to evaluate the formation of the mass-gap objects and their companions for different astrophysical scenarios. We describe our Bayesian model comparison method in Section \ref{sec:Bayes}, discuss the considered astrophysical scenarios in Section \ref{sec:populations}, present results in Section \ref{sec:results}, and conclude in Section \ref{sec:conclusion}.

\section{Bayesian model comparison} \label{sec:Bayes}

We used Bayesian model comparison to determine which astrophysical channel is a better explanation for the events GW190814 and GW200210\_092254. For each model we need to compute the posterior probability $P(A|\vec{x})$, where $A$ is the astrophysical model and $\vec{x}$ is the observed gravitational wave data from a single event. To obtain this probability, we follow \cite{2019MNRAS.486.1086M} and first consider the case in which the binary parameters $\vec{\theta}'$ are precisely known. In this case we can define the posterior probability as $P(A|\vec{\theta}')$. Using Bayes' theorem we can express this as
\begin{equation}
p(A|\vec{\theta}') = \frac{p(\vec{\theta}'|A)\pi(A)}{\pi(\vec{\theta}')}
\label{eq:Bayesknown}
\end{equation}
where $\pi(A)\propto \mathcal{R}_{A}$ is the prior probability of model $A$, which is proportional to the expected event rate density $\mathcal{R}_{A}$ and $\pi(\vec{\theta}')$ is the prior probability of $\vec{\theta}'$.  Here we have imposed uniform priors on the redshifted component masses, on the individual spin magnitudes and on the square of the luminosity distance \citep{KAGRA:2021duu}. 

The likelihood $p(\vec{\theta}'|A)$ depends on the astrophysical probability distribution $p_{\rm pop}(\vec{\theta}'|A)$ for compact object properties $\vec{\theta}'$ given by model $A$, and the detectability $p_{\rm det}(\vec{\theta}')$ of a binary with properties $\vec{\theta}'$:
\begin{equation}
p(\vec{\theta}'|A) = \frac{p_{\rm pop}(\vec{\theta}'|A)p_{\rm det}(\vec{\theta}')}{\int d\vec{\theta}p_{\rm pop}(\vec{\theta}|A)p_{\rm det}(\vec{\theta})}
\label{eq:knownlikelihood}
\end{equation}
where the denominator accounts for the fact that we only consider events that have been detected. Substituting Eq. \ref{eq:knownlikelihood} into Eq. \ref{eq:Bayesknown} gives the posterior probability we need for model comparison. $p_{\rm det}(\vec{\theta}')$ is the detection probability for an
event with true parameters $\theta'$, which we obtained by
computing the cosmic volume within which the gravitational wave with these parameters would give a signal-to-noise ratio of  $>8$ combined for two LIGO detectors with O3 sensitivity \citep{Chen:2017wpg}. 

Compared to the above case, we additionally need to take into account measurement uncertainty, i.e. that gravitational wave data provides a probability density $p(\vec{\theta}|\vec{x})$ instead of the precise value $\vec{\theta}'$. We can take this into account by marginalizing over $p(\vec{\theta}|\vec{x})$, which gives the posterior probability
\begin{equation}
    p(A|\vec{x}) = \pi(A) \frac{\int d \vec{\theta} \,p(\vec{\theta}|\vec{x})\pi(\vec{\theta})^{-1}p_{\rm det}(\vec{\theta})p_{\rm pop}(\vec{\theta}|A)}{\int d\vec{\theta} \, p_{\rm det}(\vec{\theta})\, p_{\rm pop}(\vec{\theta}|A)}.
    \label{eq:integral}
\end{equation}
Here, the distribution $p(\vec{\theta}|\vec{x})$ can be estimated using the observed data. We adopted the reconstructed $p(\vec{\theta}|\vec{x})$ for GW190814 and GW200210\_092254 from \cite{2020arXiv201014527A} and \cite{2018arXiv181112907T}, respectively.

The posterior odds favoring model $A$ over model $B$ is 
\begin{equation}
\mathcal{O}_{AB}=\frac{P(A|\vec{x})}{P(B|\vec{x})} = \frac{\mathcal{R}_{A}}{\mathcal{R}_{B}}\, \mathcal{B}
\end{equation}
Here, $\mathcal{B}$ is the Bayes factor that represent new information from observations. In general, posterior odds values $>1$ ($<1$) show that the available data prefers model $A$ ($B$), while statistically significant differentiation requires $\mathcal{O}_{AB}\gg 1$ or $\ll 1$. 

As the expected rate density of binary mergers involving different formation channels that produce objects in the lower mass gap is highly uncertain, it is currently difficult to estimate the prior odds $\mathcal{R}_{A}/\mathcal{R}_{B}$. Therefore, in the present analysis, we focus exclusively on the Bayes factor that compares models solely based on the available measurements. The ultimate determination of the origin of objects in the lower mass gap will necessitate prior expectations on rate densities, or possibly an overwhelmingly high Bayes factor from a larger number of observations.

\section{Astrophysical models} \label{sec:populations}

\begin{figure}
    \centering
\includegraphics[width=0.48\textwidth,,trim={2.2cm 0.5cm 2.5cm 0cm},clip]{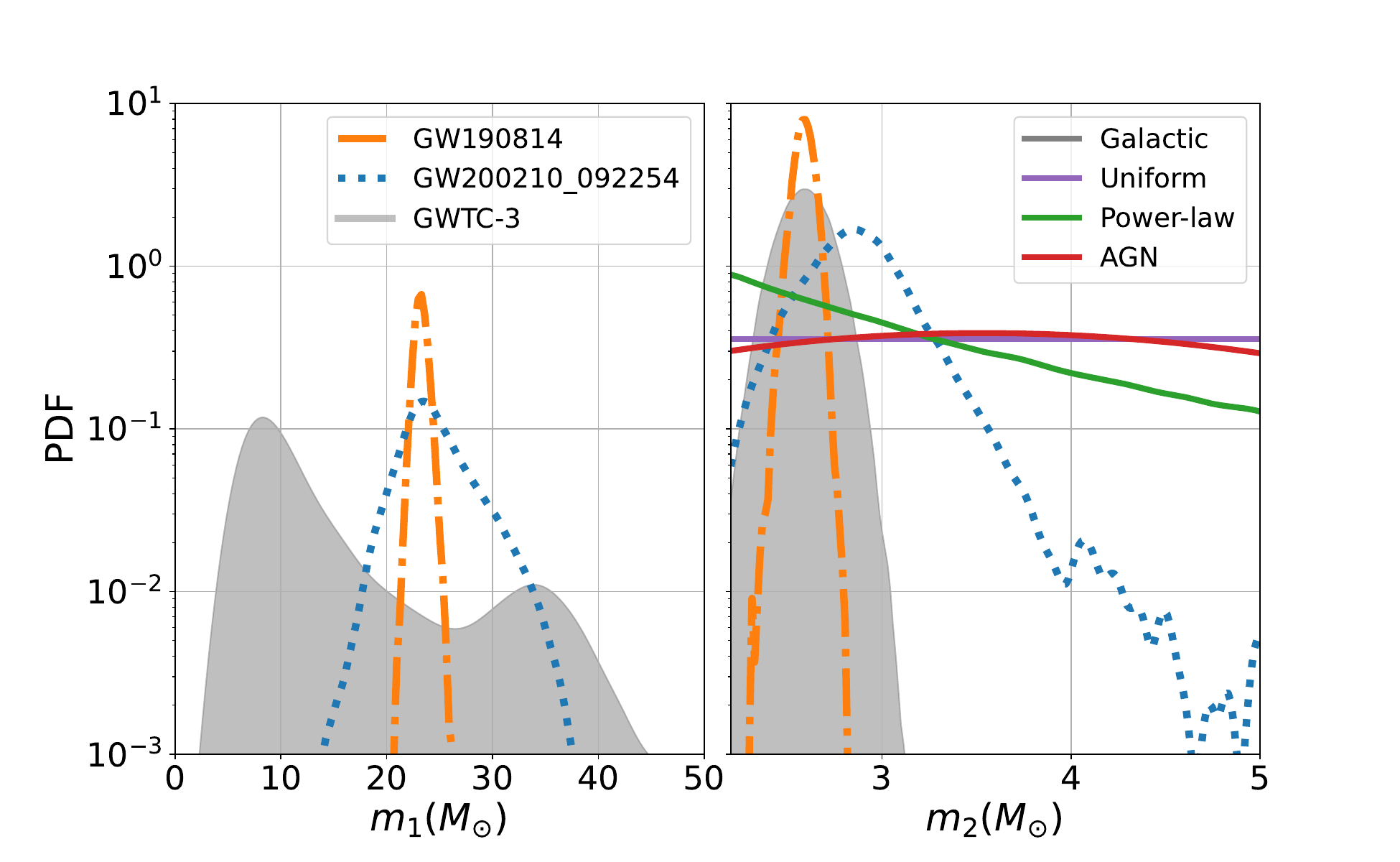}   
\caption{{\bf Probability density of primary and secondary mass distributions for GW190814 and GW200210\_092254.} Left: Primary masses in comparison to reconstructed primary mass distributions for events in the GWTC-3 catalog. Right: Secondary masses in comparison to the Galactic total binary neutron star mass distribution, along with other model distributions (see legend). }
    \label{fig:mass}
\end{figure}

We considered several characteristic astrophysical models which give different probability distributions for the expected mass distribution in the lower mass gap. We relied only on the object masses as these are the least uncertain, while we neglected spin, cosmic evolution, and other possible differences. The different mass distributions for the lower-mass gap objects, both measured and modeled, are shown in Fig. \ref{fig:mass}. We describe our four astrophysical models in the following:

\paragraph{Model 1: Binary neutron star } \, In this model we took the mass gap object to be the remnant of a binary neutron star merger. For its expected mass distribution we adopted the detected Galactic binary neutron star systems, with a total mass of $\approx 2.65\pm0.12$\,M$_\odot$ \citep{2019ApJ...876...18F}. We considered this to be the mass of the lower mass gap objects if their origin is a neutron star merger. We further incorporated the loss of $\sim0.05$\,M$_\odot$ due to mass ejection during the merger, and a loss of $\sim 0.1$\,M$_\odot$ due to gravitational wave emission (e.g., \citealt{2016PhRvD..93f4047D}). Neglecting these two processes of mass loss, however, negligibly changes our results. One caveat is that the observed binary neutron star systems in the Galaxy may not be fully representative of all binary neutron stars formed in the universe due to selection effects (see e.g. the masses $2.0^{+0.6}_{-0.3}$\,M$_\odot$ and $1.4^{+0.3}_{-0.3}$\,M$_\odot$ of the two neutron stars in merger GW190425 \citealt{2020arXiv201014527A}). However, based on this gravitational wave observation, such high-mass neutron star merger remnants are relatively rare. The results obtained in this paper are robust even when adding a small tail to the Galactic binary neutron star total mass distribution.

\paragraph{Model 2: AGNs }\, Neutron stars in AGN disks may undergo substantial accretion. AGNs also provide a conduit for bringing black holes and neutron stars closer together, resulting in mergers once the neutron stars had time to accrete from the disk.
As a fiducial AGN model we adopted the mass distribution from \cite{2020ApJ...901L..34Y} (choosing their model with maximum accretion rate of $0.7\dot{M}_{\rm Edd}$). We note that the accretion process inside AGN disks, and orbital alignment and migration within the disk, remain highly uncertain.

\paragraph{Model 3: Uniform}\, For this model we consider a uniform distribution of secondary black hole mass from $2.2$ $M_{\odot}$ to $5$ $M_{\odot}$. The motivation for this choice is that, for accretion, there may not be a particularly favored mass value such as the total mass of a binary neutron star system. Instead, accreting objects might populate the entire mass gap. Indeed, we see that this uniform distribution is similar to the AGN-assisted distribution we introduced above (see Fig. \ref{fig:mass}).

\paragraph{Model 4: Power-law }\, To model a distribution that favors lower masses, we considered a power-law mass distribution with probability density $p(m)\propto m^{-\alpha}$ with index $\alpha=2.3$, i.e. similar to the mass distribution of stars in this mass range \citep{2019NatAs...3..482K}.

\section{Results} \label{sec:results}

We have computed the Bayes factors for GW190814 and GW200210\_092254 comparing each combination of our four models discussed above in Section \ref{sec:populations}, using Bayesian model comparison described in Section \ref{sec:Bayes}. Our results are shown in Table \ref{tab:1}. We see that the neutron star merger origin (Model 1) is favored over all other models considered here. Considering the product of the Bayes factors for the two events, Model 1 is favored with a total Bayes factor of $\mathcal{B}\approx5-14$ over other models. We note that most of this statistical support comes from GW190814 because its secondary mass is far more precisely measured than is the secondary mass in GW200210\_092254.

Beyond the mass-gap objects, the primary masses in the binaries may also carry information of the mergers' origin. To qualify the similarity between the primary masses of GW190814 and GW200210\_092254, we computed the likelihood ratio of $m_1$ in GW200210\_092254 arising from the reconstructed probability density of $m_1$ in GW190814 versus arising from the reconstructed distribution of $m_1$ in all observed binary mergers in the LIGO-Virgo sample \citep{2021arXiv211103634T}. If masses from hierarchical triples that form two neutron stars and a black hole indeed result in comparable black hole masses then this ratio will tend to be higher, while if the black hole masses are quasi-random due to chance encounters then this ratio will tend to be lower. We obtained a Bayes factor of $\mathcal{B}=18$, favoring a common mass for the two triples over a wider GWTC-3-like distribution. 

\renewcommand{\arraystretch}{0.95}
\begin{table}
\begin{center}
\begin{tabular}{|l|c|c|c|}
\hline
& \bf{{GW190814}}  &  \bf{{GW200210...}} & Total\\ 
 \hline
  \hline
BNS \,-- AGN  & 8.1 & 1.68 & 13.6\\
\hline
BNS \,-- Uni. & 7.95 & 1.74 & 13.8 \\
\hline
BNS \,-- PL & 3.82 & 1.29 & 4.9 \\
\hline
AGN -- Uni. & 0.98 & 1.04 & 1.0 \\
\hline
AGN -- PL & 0.45 & 0.77 & 0.3 \\
\hline
PL \,\,\,\,\,\,-- Uni. & 2.07 & 1.35 & 2.8\\
\hline
\end{tabular}
\end{center}
\caption{{\bf Model comparison for mass gap objects.} 
Column 1 shows the two models under comparison (BNS - total mass of Galactic binary neutron stars; AGN - mass distribution expected in AGNs; Uni. - uniform mass distribution; PL - power-law mass distribution). Columns 2 and 3 represent estimated Bayes factors for GW190814 and GW200210\_092254 respectively. Column 4 shows the total Bayes factor, i.e. the product of values in columns 2 and 3.  
} 
\label{tab:1}
\end{table}

\section{Conclusion}\label{sec:conclusion}

We investigated the statistical support for a hierarchical triple origin for binary mergers GW190814 and GW200210\_092254 based on the masses of the compact objects in the binaries. Using Bayesian model comparison, we found that the  masses in the lower mass gap of these two events, which are both $\sim2.6$\,M$_\odot$ are more consistent with the total mass of binary neutron stars than other considered model distributions, with a Bayes factor $\mathcal{B}\approx 5-14$ depending on the alternative model.

We also examined the similarity of the more massive companion masses ($23.2^{+1.1}_{-1.0}$M$_\odot$ and $24.1^{+7.5}_{-4.6}$M$_\odot$) in comparison to the full $m_1$ mass distribution reconstructed based on all observed binary mergers in the GWTC-3 gravitational wave catalog. We estimate that the primary mass of GW200210\_092254 coming from the reconstructed primary mass distribution of GW190814 is favored over a random mass selection from GWTC-3 with a Bayes factor $\mathcal{B}=18$.

While our result on the secondary masses point to a hierarchical triple origin for GW190814 and GW200210\_092254 and our result on the primary mass points away from some alternative scenarios, significant caveats remain. (1) For the mass-gap objects, the total binary neutron star mass considered here is based on observed binaries in the Milky Way. These are long-lived binaries, while the total binary neutron star mass distribution might be very different if we accounted for short-lived binaries as well \citep{2020ApJ...900...13S}. In addition, if binary neutron stars are formed via chance encounters, their masses may be more widely distributed, as the neutron star mass distributions for binaries with non-neutron-star companions is wider (e.g., \citealt{2013ApJ...778...66K,2012ApJ...757...55O}). (2) While our alternative uniform and power-law models are representative of processes that allow a broader mass distribution, e.g., due to accretion or unexpected stellar evolution, actual astrophysical scenarios might result in markedly different distributions. (3) In our model comparison of the primary mass distributions, the relevant mass distribution for dynamical formation may be different from the GWTC-3 distribution (e.g., if only a small fraction of the observed binaries are of dynamical origin). (4) Our model comparison does not take into account the expected astrophysical rate densities for different scenarios. (5) The detection significance of GW200210\_092254 is limited with false alarm rate $\sim 1$\,yr$^{-1}$. These uncertainties will need to be further explored and accounted for in future work. 

Generally, however, key evidence in support or against the probes of a hierarchical triple channel described here will come from the further observations of similar events by LIGO, Virgo and KAGRA \citep{2021PTEP.2021eA101A} in current and upcoming observing runs that will make clear the emerging patterns of binary properties.

G.V. acknowledges the support of the National Science Foundation under grant PHY-2207728. 
I.B. acknowledges the support of the Alfred P. Sloan Foundation, and NSF grant PHY-2309024. 
S.R. has been supported by the Swedish Research Council (VR) under 
grant number 2020-05044, by the research environment grant
``Gravitational Radiation and Electromagnetic Astrophysical
Transients'' (GREAT) funded by the Swedish Research Council (VR) 
under Dnr 2016-06012, by the Knut and Alice Wallenberg Foundation
under grant Dnr. KAW 2019.0112,   by the Deutsche 
Forschungsgemeinschaft (DFG, German Research Foundation) under 
Germany’s Excellence Strategy – EXC 2121 ``Quantum Universe'' 
– 390833306 and by the European Research Council (ERC) Advanced 
Grant INSPIRATION under the European Union’s Horizon 2020 research 
and innovation programme (Grant agreement No. 101053985).
%
This work was performed in part at the Aspen Center for Physics, which is supported by National Science Foundation grant PHY-2210452.. Sz.M. is grateful for the generous support of Columbia University. D.V. was supported by European Research Council (ERC) under the European Union’s Horizon 2020 research and innovation programme grant agreement No 801781.  
 G.V. acknowledges
the support of the National Science Foundation under grant PHY-2207728. This material is based upon work supported by NSF's LIGO Laboratory which is a major facility fully funded by the National Science Foundation. The authors are grateful for computational resources provided by the LIGO Laboratory and supported by National Science Foundation Grants PHY-0757058 and PHY-0823459.

\bibliography{reference}{}
\bibliographystyle{aasjournal}

\end{document}